\documentclass[prl,twocolumn,superscriptaddress,showpacs]{revtex4-1}

\usepackage{amsmath,bm,amsfonts,subfigure}
\usepackage[pdftex]{graphicx}
\usepackage{epstopdf}
\usepackage{color}
\usepackage{hyperref}

\newcommand{\kinj}{k_{\mathrm{inj}}}

\begin{document}

\title{Elliptical Tracers in Two-dimensional, Homogeneous, Isotropic
Fluid Turbulence: the Statistics of Alignment, Rotation, and Nematic Order
}
\author{Anupam Gupta}
\affiliation{Center for Condensed Matter Theory,
Department of Physics, Indian Institute of Science, Bangalore 560012, India}
\author{Dario Vincenzi}
\affiliation{Universit\'e Nice Sophia Antipolis, CNRS, Laboratoire
J.A. Dieudonn\'e, UMR 7351,
06100 Nice, France}
\author{Rahul Pandit}
\altaffiliation[Also at: ]{Jawaharlal Nehru Centre for Advanced Scientific
Research, Jakkur, Bangalore, India.}
\affiliation{Center for Condensed Matter Theory,
Department of Physics, Indian Institute of Science, Bangalore 560012, India}
\date{\today}
\begin{abstract}
We study the statistical properties of orientation and rotation dynamics of
elliptical tracer particles in two-dimensional, homogeneous and isotropic
turbulence by direct numerical simulations.  We consider both the cases in
which the turbulent flow is generated by forcing at large and intermediate
length scales. We show that the two cases are qualitatively different. For
large-scale forcing, the spatial distribution of particle orientations forms
large-scale structures, which are absent for intermediate-scale forcing. The
alignment with the local directions of the flow is much weaker in the latter
case than in the former. For intermediate-scale forcing, the statistics of
rotation rates depends weakly on the Reynolds number and on the aspect ratio of
particles. In contrast with what is observed in three-dimensional turbulence,
in two dimensions the mean-square rotation rate increases as the aspect ratio
increases. 
\end{abstract}
\maketitle

The elucidation of the statistical properties of fluid turbulence
is a problem of central importance in a variety of areas that
include fluid dynamics, nonlinear dynamics, and non-equilibrium
statistical mechanics~\cite{MY75,frischbook,BJPV98,PPR09}.
Over the last decade or so, important advances have been
made in developing an understanding of the statistical properties
of homogeneous, isotropic turbulence in the Lagrangian
framework~\cite{FGV01,toschiannrev,collinsannrev}. Most of the
studies in this framework, experimental, theoretical, and
numerical, have used spherical or circular tracer particles in
three and two dimensions (3D and 2D, respectively). 
The study of the dynamics of non-spherical particles in turbulent flows
has applications in the simplest models for swimming 
micro-organisms~\cite{KS11}, ice crystals in
clouds~\cite{CL94}, and fibers in the paper
industry~\cite{LSA11}.
Recent
work in 3D turbulent flows~\cite{SK05,pumir11,toschi12,WK12,V13,CM13,GEM14} 
and in 2D low-Reynolds-number flows~\cite{WBM09-10,voth11}
has renewed interest in Lagrangian studies with 
anisotropic particles.  
We extend these studies to 2D, homogeneous and isotropic
turbulence with elliptical tracer particles.
Our study yields several interesting results, which have neither been 
obtained nor anticipated hitherto. 
We show that the 
dynamics of elliptical particles depends significantly on whether
the fluid is forced at (A) large or (B) small length scales; the 
alignment of $\bm p$, the unit vector along the semi-major axis 
of an elliptical particle, and $\nabla \times \bm \omega$, 
where $\bm \omega$ is the vorticity, is more pronounced in case (A)
than in case (B); and the statistics of the particle-rotation rate 
depends appreciably on the Reynolds number of the flow and the 
aspect ratio of the particles in case (A) but not in case (B).
Moreover, we find important
differences between the statistical properties of elliptical tracers in
2D turbulence and their counterparts for ellipsoidal particles in 
3D turbulence.
In 3D, $\bm p$ exhibits a strong alignment with
$\bm\omega$ \cite{pumir11}, the mean-square-rotation rate of $\bm p$
decreases as the aspect ratio of particles increases \cite{toschi12},
and the autocorrelation function of $\bm p$ decays exponentially,
with a correlation time increasing as a function of the Reynolds
number \cite{pumir11}. By contrast, in 2D, we show that
the alignment of $\bm p$ and $\nabla\times\bm \omega$
is much weaker than its analog in 3D, namely,
the alignment of $\bm p$ and $\bm \omega$;
in addition, 
the mean-square-rotation rate of $\bm p$ increases
as the aspect ratio of particles increases.
We thus extend significantly what is known about the differences
between 2D and 3D turbulence~\cite{frischbook,L08,PPR09,BE12}.

\begin{table*}
   \begin{tabular}{@{\extracolsep{\fill}} c c c c c c c c c c c c c c c c c}
    \hline
    Run  &$N$ & $\nu$ & $\mu$ & $f_{\mathrm{inj}}$ & $k_{\mathrm{inj}}$ & $\delta t$ & $\eta$ & $\lambda$ & $Re_{\lambda}$ &
$T_{\mathrm{eddy}}$ & $\tau_{\eta}$ & $\tau_{S_{11}}$ & $\tau_{S_{12}}$ &
$\tau_{\omega}$ & $N_p$\\
   \hline
    {\tt A1}  & $2048$ & $5\times10^{-5}$ & $0.01$ & $1.9\times 10^{-6}$ & $2$ & $4.0\times 10^{-3}$  & $3.3 \times 10^{-2}$	&
$0.460$ & $199$ & $23.5$ & $21.3$ & $92.2$ & $94.4$
& $46.6$ & $10^4$\\
\hline
    {\tt B1}  & $2048$ & $5\times10^{-5}$ & $0.01$ & $7.8\times 10^{-3}$ & $50$ & $1.0\times 10^{-3}$  & $3.8 \times 10^{-3}$	&
$0.053$ & $202$ & $1.34$ & $0.28$ & $1.59\times 10^{-2}$ & $1.61\times 10^{-2}$
& $8.01\times 10^{-3}$ & $10^4$\\
\hline
    {\tt A2}  & $2048$ & $5\times10^{-5}$ & $0.01$ & $8\times 10^{-6}$ & $2$ &  $2.0\times 10^{-3}$  & $2.6 \times 10^{-2}$	&
$0.470$ & $382$ & $14.7$ & $12.0$ & $24.7$ & $24.1$ & $12.2$
& $10^4$\\
\hline
{\tt B2}  & $2048$ & $5\times10^{-5}$ & $0.01$ & $1.65\times 10^{-2}$ & $50$ &  $5.0\times 10^{-4}$ & $3.2 \times 10^{-3}$    &
$0.063$ & $385$ & $1.37$ & $0.21$ & $8.66\times 10^{-3}$ & 
$8.73\times 10^{-3}$ & $4.35\times 10^{-3}$ &  $10^4$\\
\hline
    {\tt A3}  & $2048$ & $5\times10^{-5}$ & $0.01$ & $1.5\times 10^{-5}$ & $2$ &  $2.0\times 10^{-3}$  & $2.0 \times 10^{-2}$	&
$0.474$ & $536$ & $11.1$ & $8.38$ & $12.6$ & $9.98$ & $5.55$
& $10^4$\\
\hline
    {\tt B3}  & $2048$ & $5\times10^{-5}$ & $0.01$ & $2.5\times 10^{-2}$ & $50$ &  $5.0\times 10^{-4}$  & $3.0 \times 10^{-3}$	&
$0.069$ & $539$ & $1.17$ & $0.17$ & $1.55\times 10^{-3}$ & $1.56\times 10^{-3}$
& $7.79\times 10^{-4}$ & $10^4$\\
\hline
\end{tabular}
\caption{\small
The parameters for our DNS runs. 
Here, $\eta \equiv(\nu^3/\varepsilon)^{1/4}$ is the dissipation scale, 
$\lambda \equiv \sqrt {\nu \mathcal{E}/\varepsilon}$ the Taylor-microscale,
$Re_{\lambda} = u_{rms} \lambda/\nu$ the Taylor-microscale Reynolds number, $T_{\mathrm{eddy}} \equiv  \Sigma_k \frac {(E(k)/k)}{E(k)}/u_{rms}$
the eddy-turn-over time, and  $\tau_{\eta} \equiv \sqrt{\nu/\varepsilon}$ the Kolmogorov
time scale, where 
${\mathcal E} \equiv \langle \frac 1 2 |{\bm u(\bm x},t)|^2\rangle_{{\bm x},t}$, where $\langle\cdot\rangle_{{\bm x},t}$
denotes a space-time average, is the total kinetic energy of the flow,
$u_{rms}$ is the root-mean-square velocity, $\varepsilon$ is the kinetic-energy dissipation rate, and
$E(k)\equiv\sum_{k-1/2< k' \le k+1/2} k'^2 \langle |\widetilde{
\psi}({\bm k'},t)|^2 \rangle _t$, where $\langle\cdot\rangle_t$ indicates a time average over the statistically steady state, is the fluid-energy
spectrum. The Lagrangian
correlation times of $S_{11}$, $S_{12}$, and $\omega$ are defined as
$\tau_{S_{11}}\equiv\langle S_{11}^2\rangle^{-1}\int_0^\infty\langle S_{11}(t)
S_{11}(0)\rangle\, dt$,
$\tau_{S_{12}}\equiv\langle S_{12}^2\rangle^{-1}\int_0^\infty\langle S_{12}(t)
S_{12}(0)\rangle\, dt$,
 and 
$\tau_\omega\equiv\langle \omega^2\rangle^{-1}\int_0^\infty\langle \omega(t)
\omega(0)\rangle\, dt$, respectively.} 
\label{table:para}
\end{table*} 

The 2D, incompressible Navier--Stokes equations can be written
in terms of the stream-function $\psi$ and the 
vorticity $\bm{\omega} = \nabla \times {\bm u}({\bm x},t) \equiv \omega {\hat {\bm z}}$, 
where ${\bm u}\equiv(-\partial_y \psi, \partial_x \psi)$ is the fluid 
velocity at the point ${\bm x}$ and time $t$, and $\hat {\bm z}$ is the unit
normal to the fluid film:
\begin{eqnarray}
D_t{\omega} = \nu \nabla^2 {\omega}
	      - \mu \omega   + f_{\omega};
~~~~
\nabla^2 {\psi}  = {\omega}.
						\label{ns}
\end{eqnarray}
Here $D_t\equiv\partial_t + {\bm u}\cdot\nabla$, the uniform solvent density $\rho
= 1$, $\mu$ is the coefficient of friction
(which is always present in experimental fluid films~\cite{PP10}),
and $\nu$ is the kinematic viscosity of
the fluid. We use a zero-mean, Gaussian stochastic forcing with
$\langle\tilde{f}_\omega(\bm k) \tilde{f}_\omega(\bm k')\rangle= 
A(\bm k)\delta(\bm k+\bm k'),$
where $A(\bm k)=f_{\mathrm{inj}}$ if $\vert\bm k\vert=k_{\mathrm{inj}}$ and zero otherwise,
the tilde denotes a spatial Fourier transform, and $k_{\mathrm{inj}}$ is
the length of
the energy-injection wave vector. 
The configuration of an elliptical particle is given by the position
of its center of mass, $\bm x_c$, and by the axial unit vector
$\bm p=(\cos\theta,\sin\theta)$ 
that specifies the orientation of the semi-major axis with respect to a fixed direction.
The elliptical particles we consider are neutrally buoyant, of uniform composition,
and much smaller than the viscous dissipation scale, so the velocity gradient
is uniform over the size of a particle. 
In addition, we study
suspensions that are sufficiently dilute for  
hydrodynamic particle--particle interactions to be disregarded.
Under the above assumptions, $\bm x_c$ satisfies
\begin{eqnarray}
\dot{\bm x}_c = {\bm u}({\bm x}_c(t),t) ;
            \label{lagrangianCM}
\end{eqnarray}
and the time evolution of the orientation is given by the 
Jeffery equation~\cite{J22}, which reduces in a 2D, incompressible
flow to the following one for the angle $\theta$:
\begin{equation}
\dot{\theta} = \textstyle\frac{1}{2}\omega + \gamma(\alpha)[\sin(2\theta)
S_{11}-\cos(2\theta) S_{12}],
\label{jefferytheta}
\end{equation}
where $S_{ij}=(\partial_i u_j+\partial_j u_i)/2$ are the components of the
rate-of-strain tensor evaluated at $\bm x_c$,
$\gamma(\alpha) \equiv(\alpha^2-1)/(\alpha^2+1)$, and
$\alpha$ is the 
ratio of the lengths of the semi-major and semi-minor axes of the elliptical 
particle; $\gamma$ varies from 0 (circular disks) to 1 (thin rods).

Our direct numerical simulation (DNS) of Eqs.~(\ref{ns})-(\ref{jefferytheta}) uses periodic 
boundary conditions over a square domain with side $ L = 2 \pi$, 
a pseudospectral method~\cite{canuto88} with $N^2=2048^2$ 
collocation points, the $2/3$ dealiasing rule, and, for the time evolution,
a second-order, exponential-time-differencing Runge--Kutta 
method~\cite{cox02,PRMP11}. 
For the integration of Eq.~(\ref{lagrangianCM}),
we use an Euler scheme, because, in one  time step $\delta t$, 
a tracer particle crosses roughly one-tenth of a grid spacing. 
At off-lattice points, we evaluate the particle velocity from 
the Eulerian velocity field by using a bilinear-interpolation 
scheme~\cite{num_recp}. 
Finally, we integrate Eq.~(\ref{jefferytheta}) by using an Euler scheme,
with the same time step as for Eq.~\eqref{lagrangianCM}; and, 
at $t=0$, the orientation angles are uniformly distributed
over $[0,2\pi]$.
We track $N_p=10^4$ particles over time to obtain the statistics of
particle alignment and rotation for $12$ different values of
$0 < \gamma \leq 1$.
We collect data for averages only when our system has reached a non-equilibrium statistically steady state,
\textit{i.e.}, for $t>20T_{\mathrm{eddy}}$, where $T_{\mathrm{eddy}}$ is the
integral-scale eddy-turn-over time of the flow. 
The parameters used in our simulations are given in Table~\ref{table:para}.
Our study consists of two sets of simulations (A) and (B) 
at comparable values of $Re_{\lambda}$, the Taylor-microscale
Reynolds numbers. In (A), the flow is
forced at small $\kinj$ (\textit{i.e.}, a large length scale); 
in (B), it is forced at an intermediate value of $\kinj$ 
(\textit{i.e.}, an intermediate length scale);
even in case (B) $\kinj$ is small enough that the energy 
spectrum displays both a part with an inverse-energy cascade and a part
with a forward cascade of enstrophy.
We have also performed simulations at a lower resolution ($N=1024$) and obtained
similar results, so our study do not suffer from finite-resolution effects.

\begin{figure}
\includegraphics[width=\columnwidth]{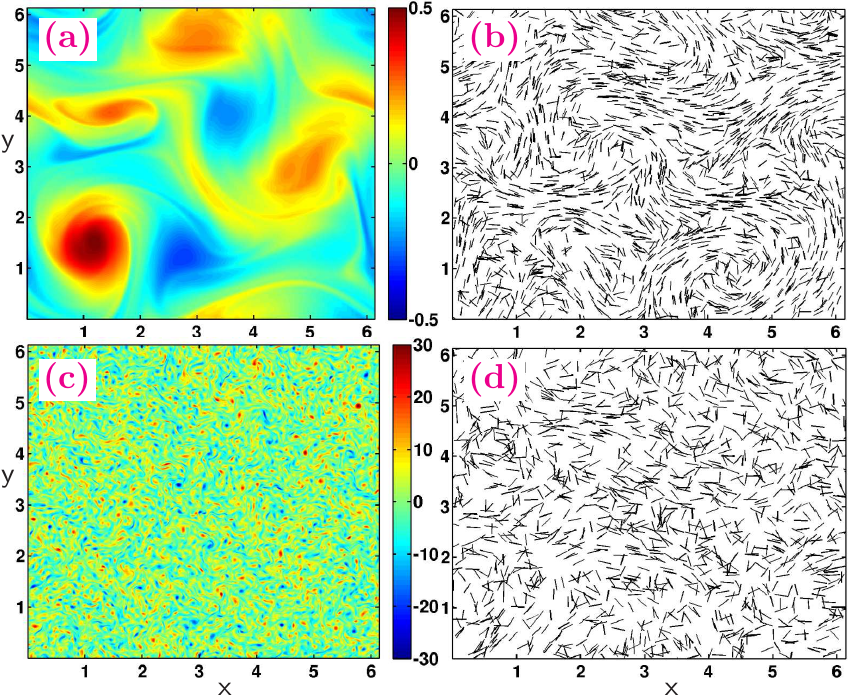}
\caption {(Color online) (a) Pseudocolor plot of $\omega$ at $t=17.5$ for run 
\texttt{A3}; (b) particle positions and
orientations at $t=17.5$ for run \texttt{A3} and $\gamma=1$;
(c) pseudocolor plot of $\omega$ at $t=17.5$ for run \texttt{B3}; 
(d) particle positions and
orientations at $t=17.5$ for run \texttt{B3} and $\gamma=1$. 
The number of particles shown in (b) and (d) is $2\times 10^3$.
For the spatiotemporal evolution of these plots see Ref.~\cite{SM}.}
\label{snapshot}
\end{figure}
In Fig.~\ref{snapshot}(a), 
we show a  pseudocolor plot of $\omega$ for case (A)
at a representative time in the statistically steady state; and Fig.~\ref{snapshot}(b) shows
the positions and the orientations of particles at the same time;
an elliptical particle is represented here by a
black line whose center indicates $\bm x_c$ and 
whose orientation is that of $\bm p$.
Analogous plots for case (B) are given in Figs.~\ref{snapshot}(c) 
and~\ref{snapshot}(d).
Figure~\ref{snapshot} suggests
that the particle dynamics is qualitatively different
in cases (A) and (B). In particular, in the former case, 
the orientation of particles is such that we see large-scale
structures, which are absent in the latter case. To quantify this behavior, we study the statistics of the alignment of
particles with the local directions of the flow.

\begin{figure}
\includegraphics[width=0.5\columnwidth]{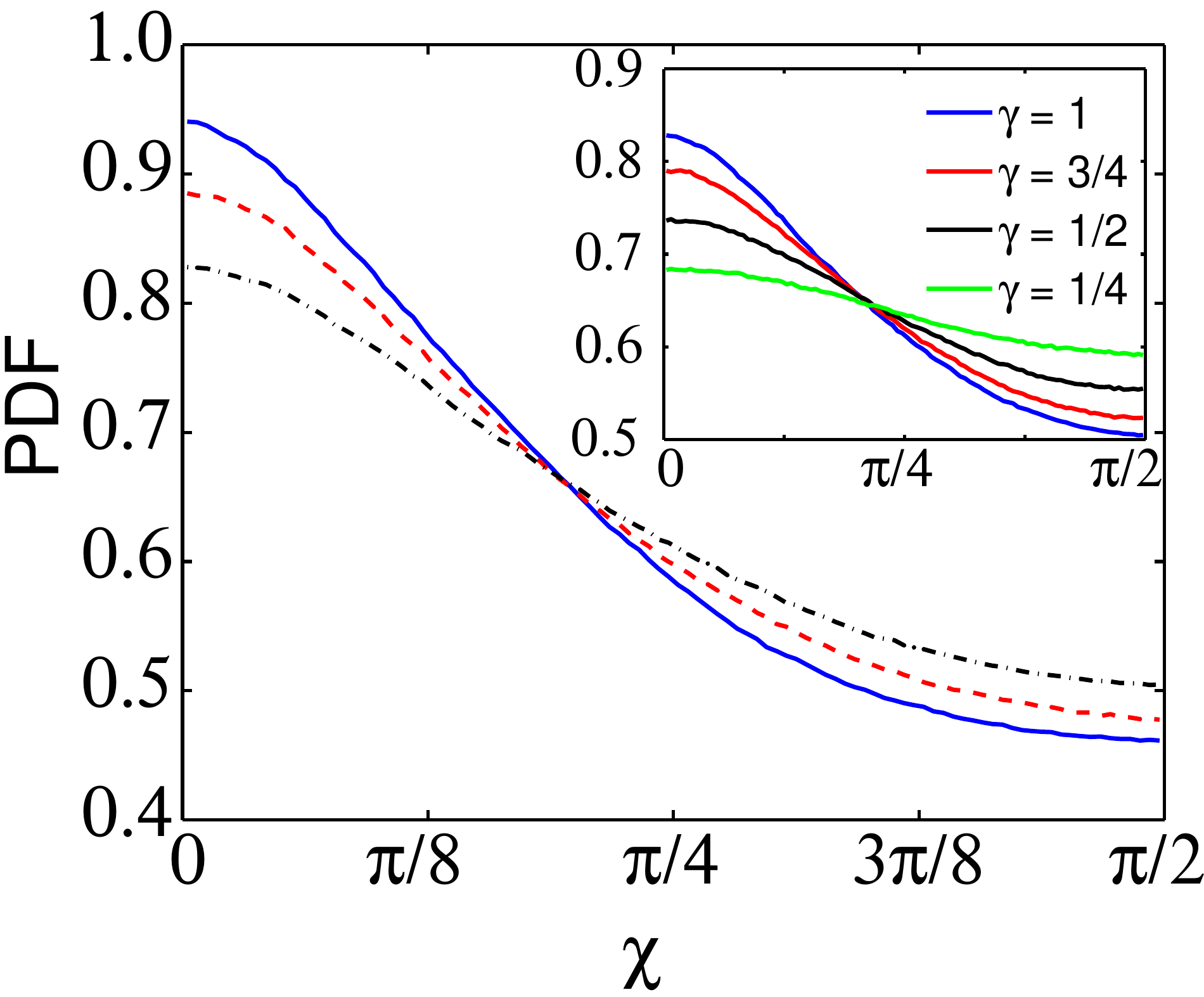}%
\hfill%
\includegraphics[width=0.5\columnwidth]{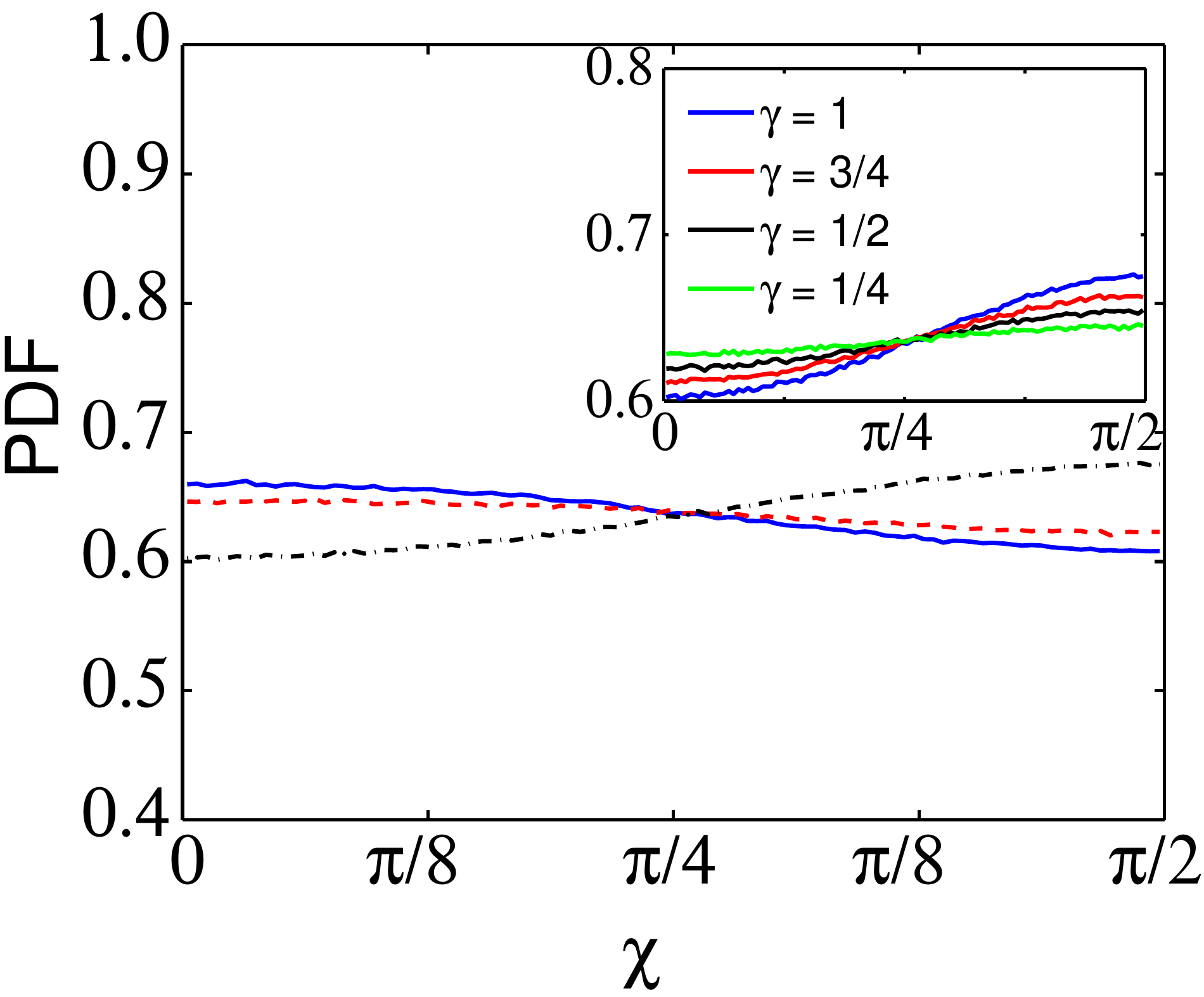}
\caption{(Color online) PDFs of the angle between 
$\bm p$ and $\bm\nabla\times\bm\omega$. Left: run \texttt{A1} (solid, blue 
curve), run \texttt{A2} (dashed, red curve), run \texttt{A3} (dot-dashed, 
black curve) for $\gamma=1$. The inset shows the same PDF for run \texttt{A3} 
and different values of $\gamma$ (from top to bottom: $\gamma=1,3/4,1/2,1/4$.)
Right: run \texttt{B1} (solid, blue curve),
run \texttt{B2} (dashed, red curve), run \texttt{B3} (dot-dashed, black curve) 
for $\gamma=1$. The inset shows the same PDF for run \texttt{B3} and different
values of $\gamma$ (from top to bottom: $\gamma=1,3/4,1/2,1/4$.)}
\label{fig:curl}
\end{figure}
The curl of the vorticity is tangent to the isolines of $\omega$; a strong 
alignment between $\bm p$ and
$\nabla\times\bm\omega$ would thus indicate a significant correlation between
the spatial distribution of particle orientations and the vorticity field. 
Figure~\ref{fig:curl} shows the probability density function
(PDF) of the angle $\chi$ between $\bm p$ and $\nabla\times\bm\omega$.
In case (A), $\bm p$ tends to align
with $\nabla\times\bm\omega$, but the alignment is not very strong.
A careful inspection of Fig.~\ref{snapshot} shows indeed that 
the spatial distribution of particle orientations does not closely reproduce
the isolines of vorticity. Moreover,
the alignment weakens as $\gamma$ decreases
and $Re_\lambda$ increases (Fig.~\ref{fig:curl}).
In case (B), the PDF of $\chi$ depends very weakly on $\chi$, \textit{i.e.}, the elliptical tracers do not
exhibit a definite preferential orientation with respect to $\nabla\times\bm\omega$. We
observe this behavior for
all the values of $Re_\lambda$ considered in Fig.~\ref{fig:curl}.

An examination of the
statistics of $\chi$ shows the first, remarkable difference between 
the dynamics of non-spherical tracers in 3D and 
that of elliptical particles in 2D.
In 3D, the tracer particles align strongly with $\bm\omega$~\cite{pumir11}. 
This behavior has been explained in 
Ref.~\cite{pumir11} by arguing that, if viscosity is disregarded,
the equation describing the Lagrangian evolution of $\bm \omega/\vert\bm\omega\vert$ is equivalent to
the evolution equation for the axial unit vector of a thin rod.
In 2D, an analogous equivalence exists, because
$(\nabla\times\bm\omega)/\vert\nabla\times\bm\omega\vert$ satisfies the Jeffery 
equation with $\gamma=1$ (provided that $\nu=0$).
In 2D, this formal equivalence does not yield
a strong alignment between $\bm p$ and $\nabla\times\bm\omega$
because the effect of the viscosity
on $\nabla\times\bm\omega$ in 2D
is more important than its effect
on $\bm\omega$ in 3D.
The aforementioned equivalence also explains why the alignment of particles
with $\nabla\times\bm\omega$ becomes weaker as their aspect ratio decreases;
and indeed the evolution equation for~$\bm p$ increasingly deviates from that
for $(\nabla\times\bm\omega)/\vert\nabla\times\bm\omega\vert$.
The decrease of the probability of
alignment with increasing $Re_\lambda$
is, on the contrary, attributable to the increase 
of the fluctuations of the components of $\nabla\bm u$.

\begin{figure}
\includegraphics[width=.5\columnwidth]{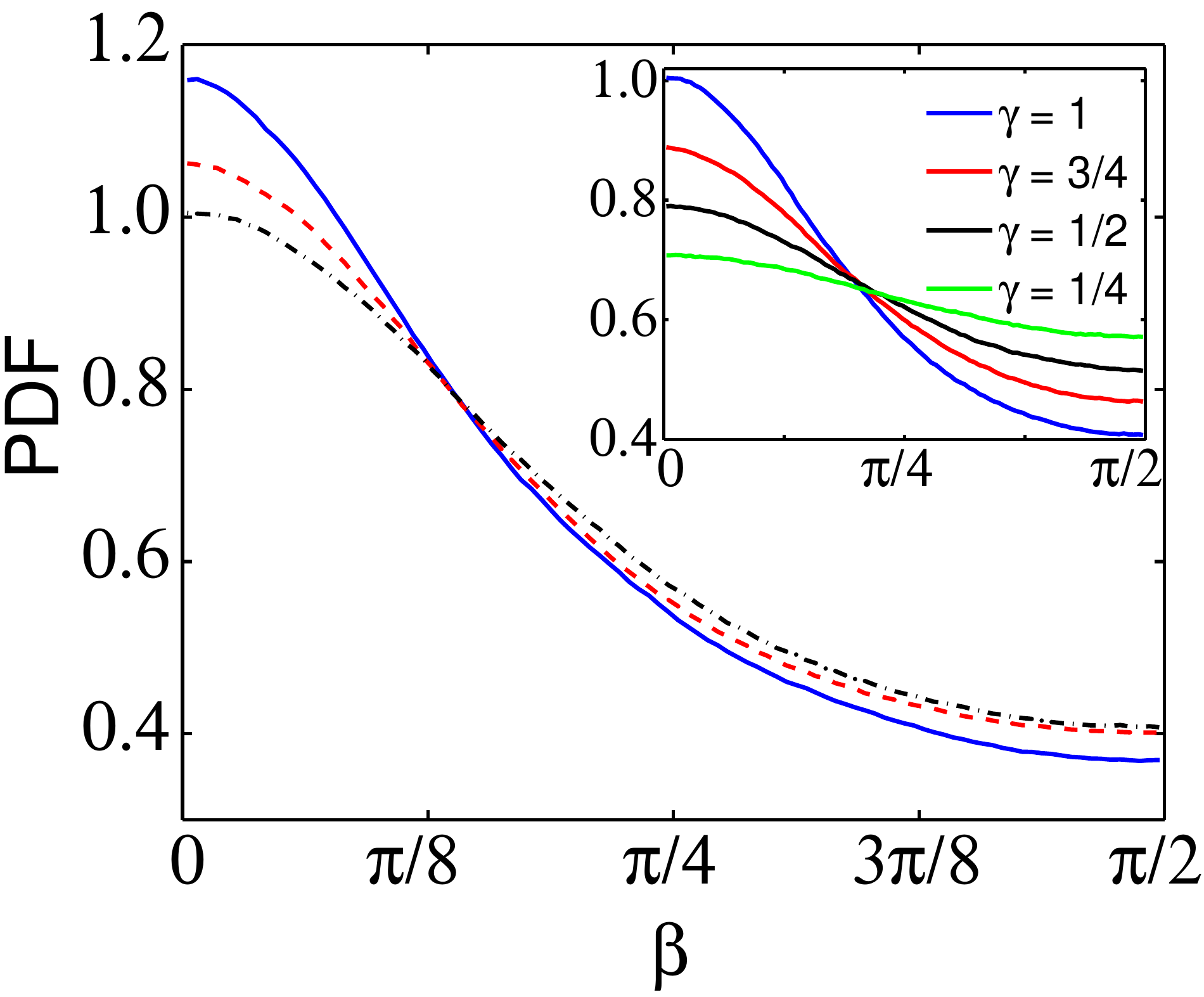}%
\hfill%
\includegraphics[width=.5\columnwidth]{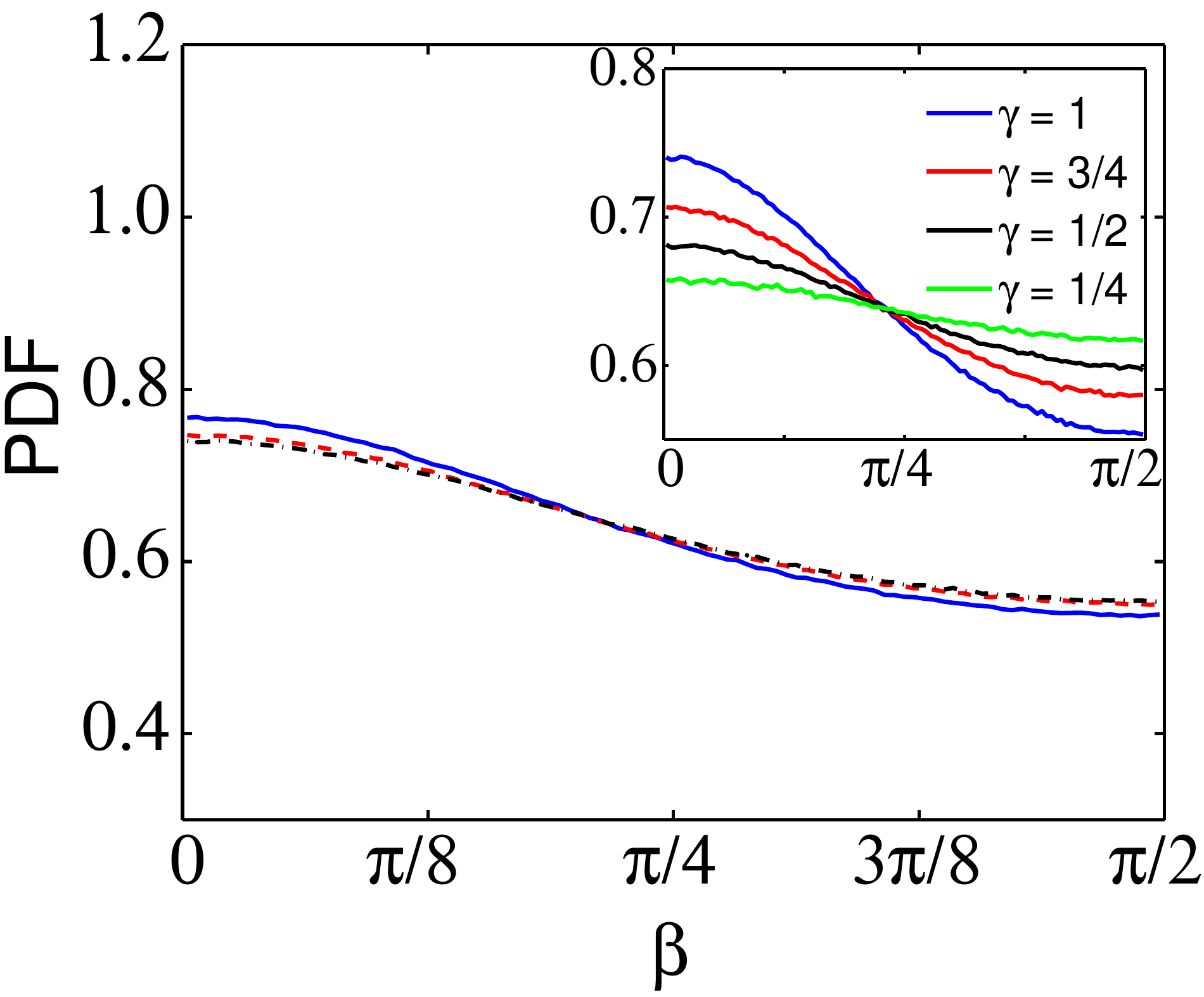}%
\caption{(Color online) PDFs of the angle $\beta$ 
between $\bm p$ and $\bm e_1$
for $\gamma=1$ and different $\mathit{Re}_\lambda$ in case (A)
(left) and in case (B) (right).
The insets show the PDF of $\beta$ for different values of $\gamma$
for runs \texttt{A3} (left) and \texttt{B3} (right).
The color code is the same as in Fig.~\ref{fig:curl}.}
\label{fig:e_1}
\end{figure}

The eigenvectors of the $S$ form a Lagrangian 
orthogonal frame of reference.
In Fig.~\ref{fig:e_1}, we show the
PDF of the angle $\beta$
between $\bm p$ and the eigenvector $\bm e_1$, associated with the positive eigenvalue
of $S$.
Particles tend to align with $\bm e_1$, but the alignment is weaker 
in case (B) than in case (A). The alignment becomes weaker as
$\gamma$ decreases, because the contribution of $S$ to 
the evolution of $\bm p$ diminishes [see Eq.~\eqref{jefferytheta}].
The tendency of particles to align with $\bm e_1$ diminishes 
as $\mathit{Re}_\lambda$ increases,
\textit{i.e.}, as turbulent fluctuations are enhanced.
The moderate degree of alignment, shown in Fig.~\ref{fig:e_1}, is comparable with that found for rods 
in 2D, low-Reynolds-number flows~\cite{voth11} and in 3D,
homogeneous, isotropic turbulence~\cite{pumir11}.

We have calculated the conditional PDFs of the alignment of particles conditioned on the
sign of the Okubo--Weiss parameter~\cite{OW,PRMP11}, which distinguishes
between vortical and extensional regions of the flow; the conditional PDFs
do not deviate from their unconditional counterparts~\cite{SM}.

\begin{figure}
\includegraphics[width=0.5\columnwidth]{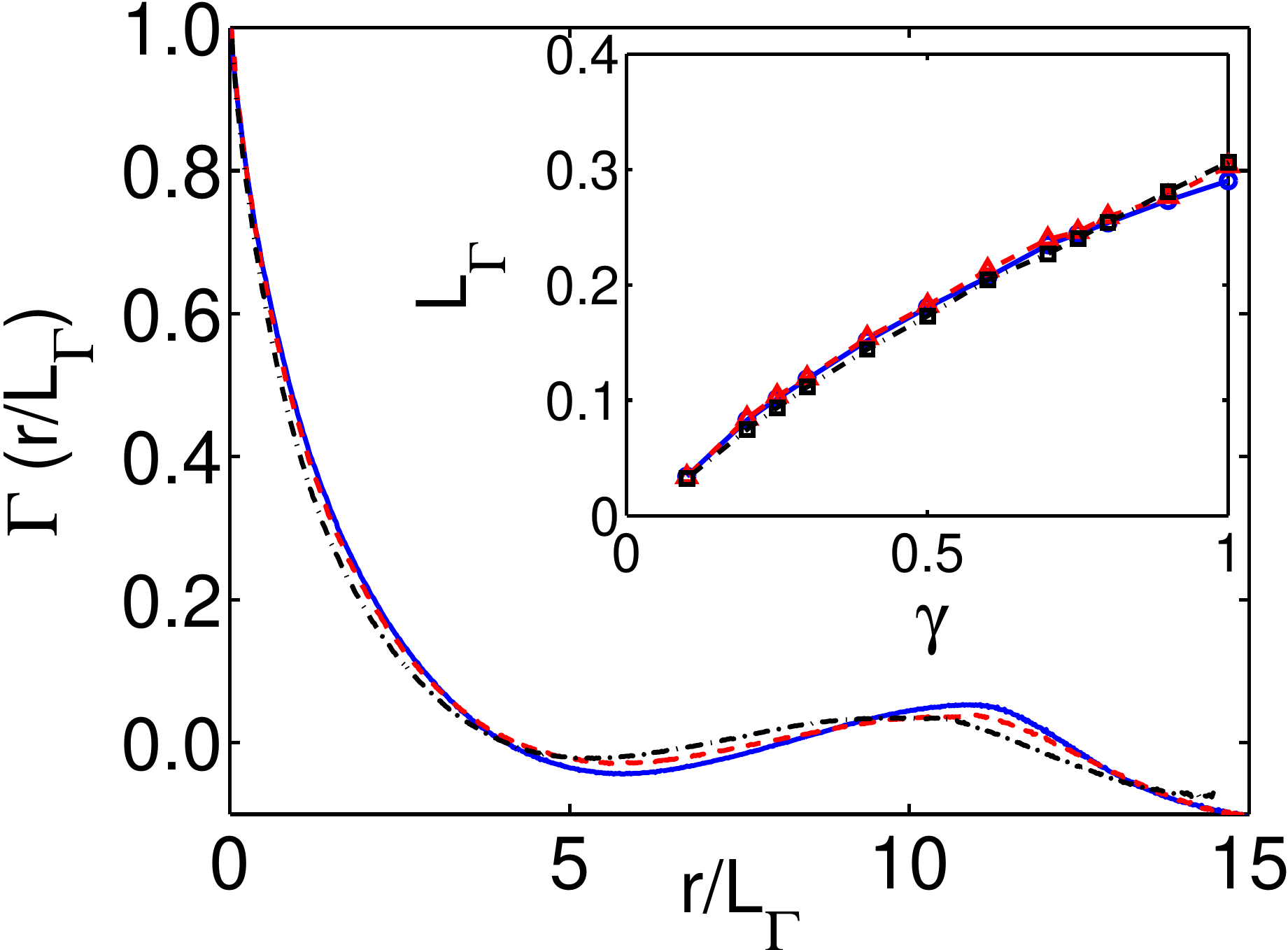}%
\hfill%
\includegraphics[width=0.5\columnwidth]{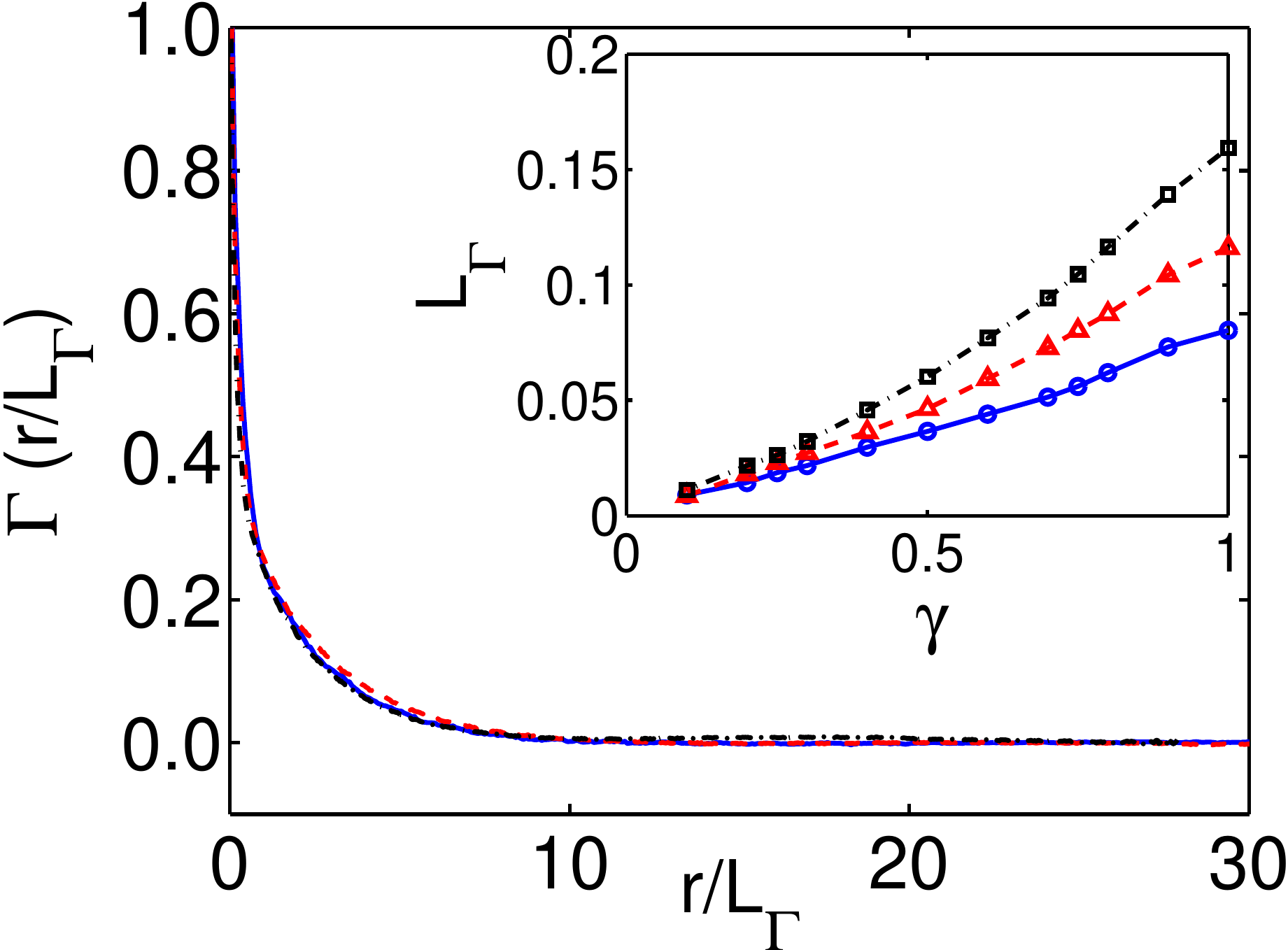}
\caption{(Color online) Single-time two-point
correlation function of $M$
as a function
of the space separation rescaled by the correlation length.
Left: run \texttt{A1} (solid, blue curve),
run \texttt{A2} (dashed, red curve), run \texttt{A3} (dot-dashed, black curve) for $\gamma=1$. 
Right: run \texttt{B1} (solid, blue curve),
run \texttt{B2} (dashed, red curve), run \texttt{B3} (dot-dashed, black curve) for $\gamma=1$. 
The insets show the correlation length as a function of
$\gamma$; the color code is the same as in the main plots.}
\label{fig:nematic}
\end{figure}

To quantify the spatial distribution of particle orientations, we define
the correlation function 
$\Gamma (r) = [\langle M(\bm r,t) M(\bm 0,t) \rangle
- \langle M(\bm r,t) \rangle \langle M(\bm 0,t) \rangle]/\langle M^2\rangle$,
where $ M(\bm r,t) \equiv (2 \cos^2\theta(\bm r,t) - 1)$
is the local nematic order parameter in 2D~\cite{CL95} and 
$\langle\cdot\rangle$ denotes an average over time and over the tracer particles.
The function $\Gamma(r)$ is shown in
Fig.~\ref{fig:nematic} for different values of~$Re_\lambda$. 
In both the cases (A) and (B), 
the shape of $\Gamma(r)$
depends only weakly on $\mathit{Re}_\lambda$.
However, in case (A), the order parameter of rods is correlated up to distances
of the order of $5\%$ $L$ and is anti-correlated
at large $r$; in case (B), $\Gamma(r)$ decays
exponentially to zero. These behaviors are in accordance with the spatial distributions of
orientations shown in Fig.~\ref{snapshot}.
Furthermore, in case (A), the correlation length $L_\Gamma=[\int_0^\pi 
\Gamma(r)dr]/\Gamma(0)$
depends weakly on $\mathit{Re}_\lambda$, because the value of $L_\Gamma$ is 
determined principally by $k_\mathrm{inj}$; in case (B), 
the size of large-scale flow structures increases
with increasing $\mathit{Re}_\lambda$~\cite{BM10}; 
hence, $L_\Gamma$ increases accordingly.
In both cases, $L_\Gamma$
is obviously an increasing function of $\gamma$.

\begin{figure}
\includegraphics[width=0.5\columnwidth]{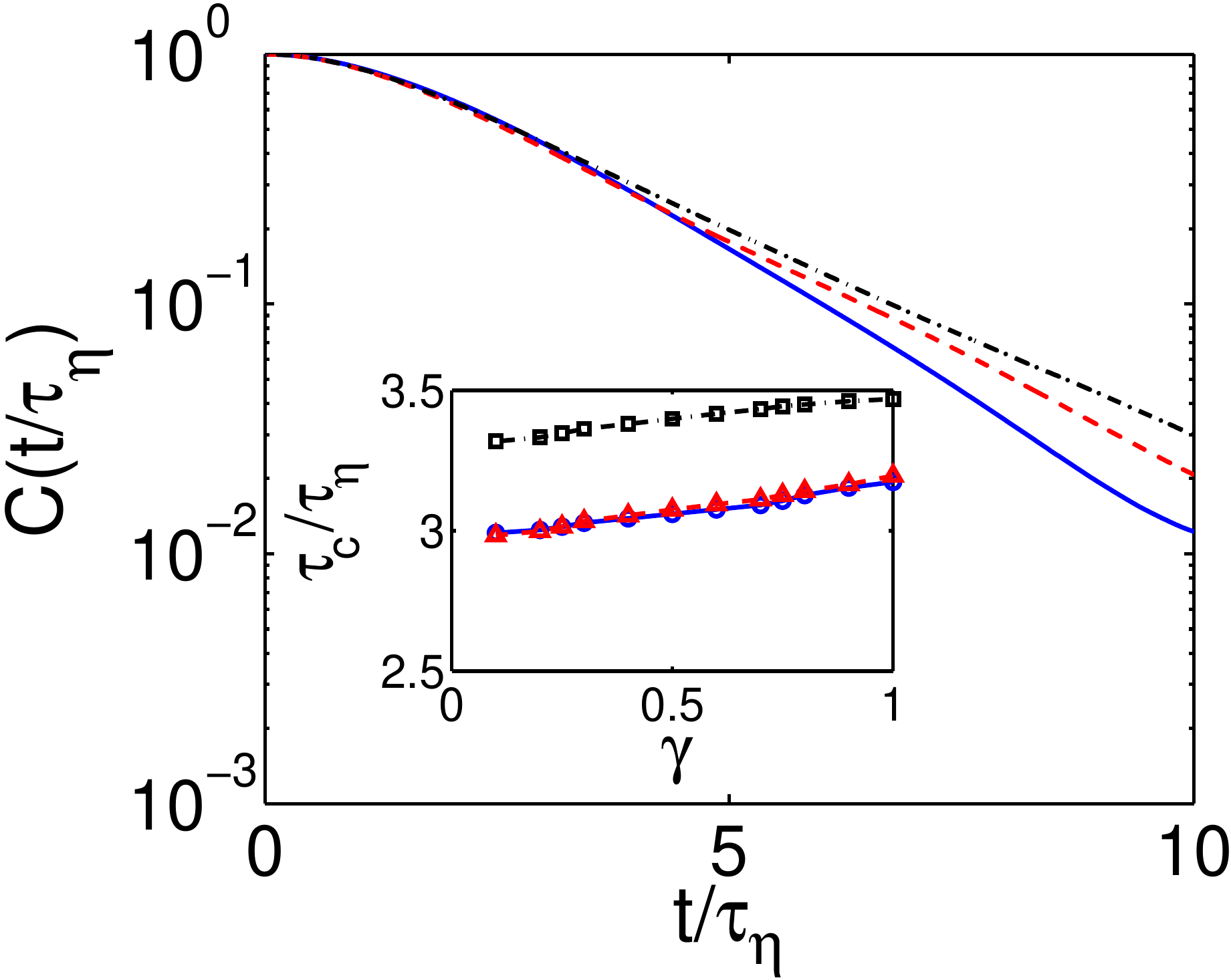}%
\hfill%
\includegraphics[width=0.5\columnwidth]{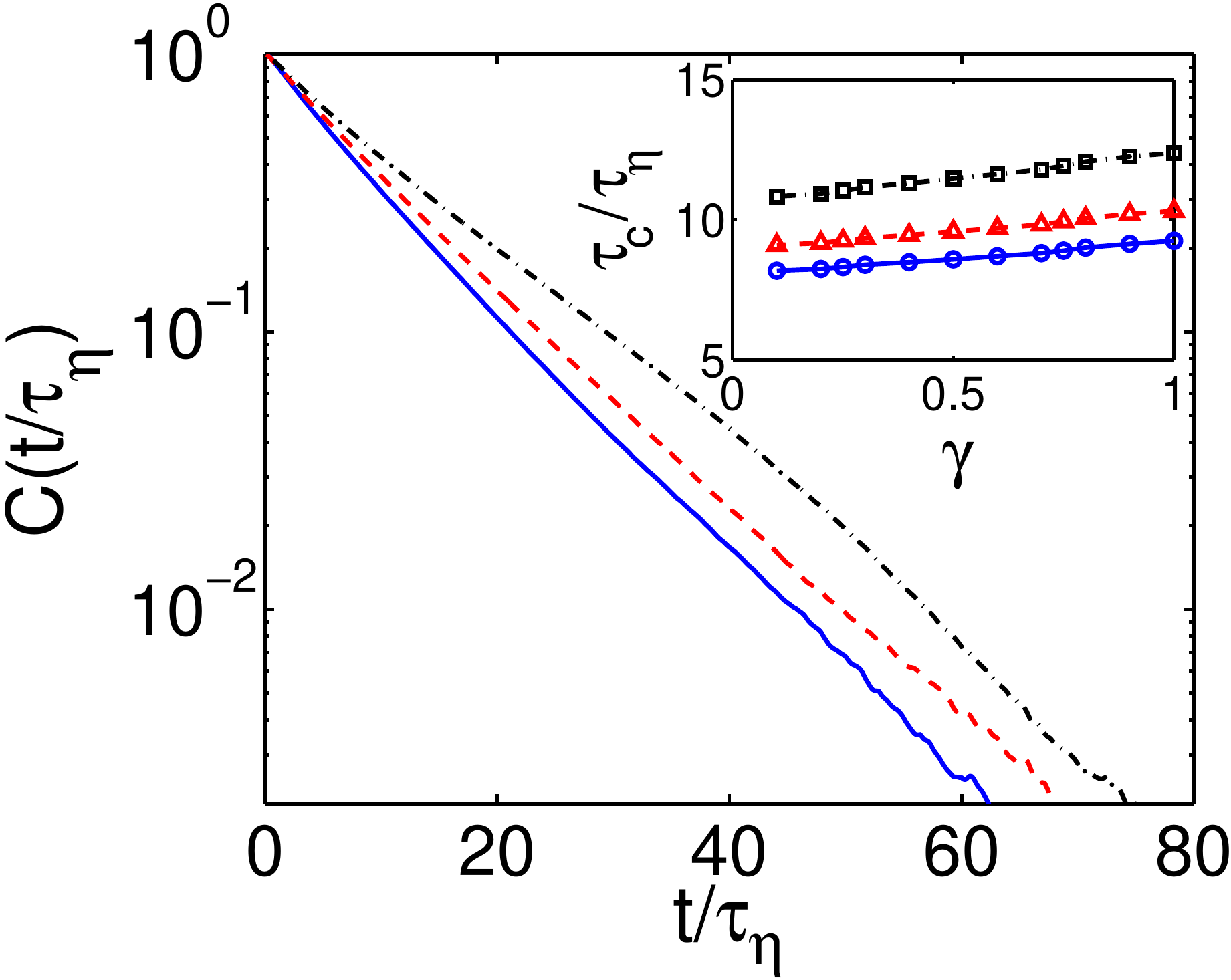}
\caption{(Color online) Autocorrelation function of $\bm p$ as a function
of the time separation rescaled by $\tau_\eta$ for different 
$\mathit{Re}_\lambda$ in case (A) (left) and in case (B) (right).
The insets show the correlation time rescaled by $\tau_{\eta}$ 
as a function of $\gamma$ for different $\mathit{Re}_\lambda$. The color 
codes are the same as in Fig.~\ref{fig:nematic}.}
\label{fig:autocorr}
\end{figure}

Let us now examine the temporal autocorrelation function of $\bm p$.
Both in cases (A) and (B), $C(t)=\langle\bm p(t)\cdot\bm p(0)\rangle$
decays exponentially to zero (Fig.~\ref{fig:autocorr}), but 
the correlation time  $\tau_c=\int_0^\infty C(t)dt$ is much shorter in the
former case. The ratio $\tau_c/\tau_{\eta}$ increases 
as a function of both $Re_\lambda$ and $\gamma$;
this behavior is similar to that observed in 3D turbulence,
where the orientational dynamics of spheres decorrelates faster than that
of rods~\cite{pumir11}.

\begin{figure}
\includegraphics[width=0.495\columnwidth]{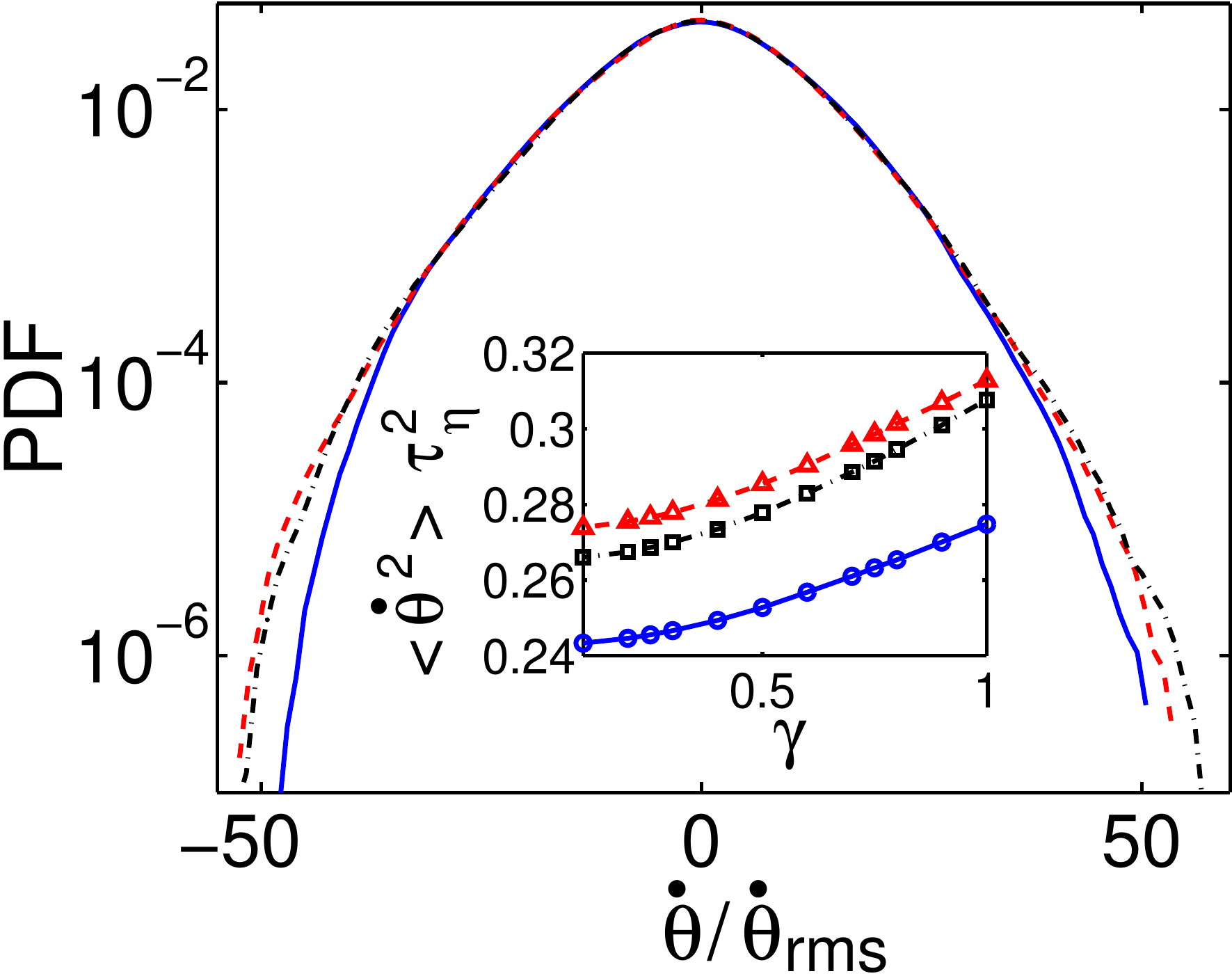}%
\hfill%
\includegraphics[width=0.495\columnwidth]{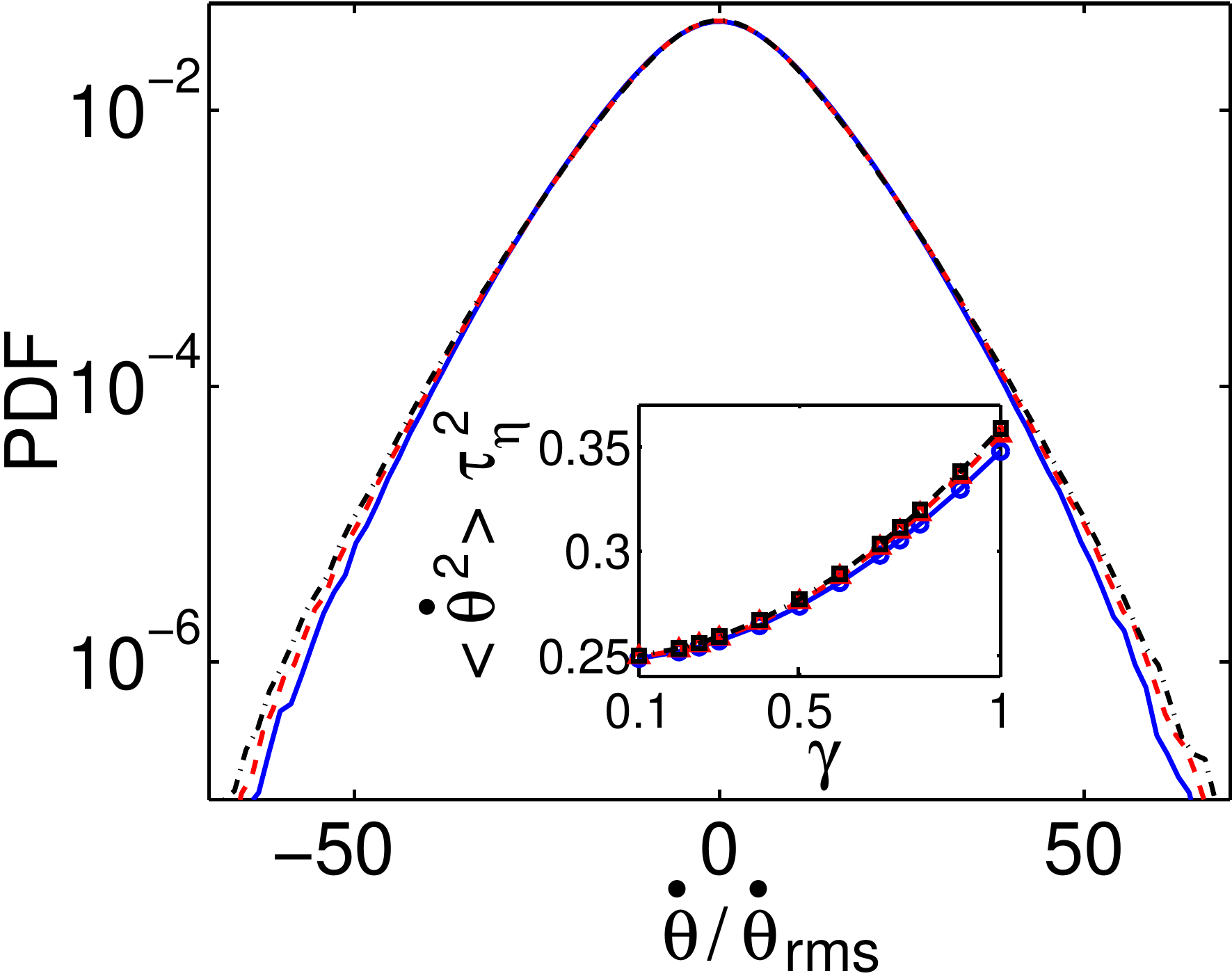}
\caption{(Color online) PDFs of $\dot{\theta}$
rescaled by
$\dot{\theta}_{\mathrm{rms}}=\sqrt{\langle\dot{\theta}^2\rangle}$
for $\gamma=1$ and different $\mathit{Re}_\lambda$
in case (A) (left) and in case (B) (right).
The insets show
the mean-square rotation rate multiplied by $\tau_\eta^2$ as a function of $\gamma$. The color codes are the same
as in Fig.~\ref{fig:nematic}.}
\label{fig:pdf-rate-Re}
\end{figure}

Figure~\ref{fig:pdf-rate-Re} 
shows the PDFs 
of the rotation rate $\dot{\theta}$ of particles
for different values of $\mathit{Re}_\lambda$ 
(for the analogous PDFs at fixed $\mathit{Re}_\lambda$  
and different $\gamma$, see~\cite{SM}).
Very large fluctuations characterize the statistics
of $\dot{\theta}$, as has been 
observed in 3D turbulence~\cite{toschi12}. 
However, the probability of large fluctuations increases with increasing $\gamma$ and $\mathit{Re}_\lambda$
in case (A), whereas it depends weakly
on $\gamma$ and $\mathit{Re}_\lambda$ in case (B).
The main difference between 2D and 3D is the dependence of the mean-squared-rotation rate
$\langle\dot{\theta}^2\rangle$ upon $\gamma$.
In 3D, $\langle\dot{\theta}^2\rangle$ decreases as $\gamma$ increases and is thus smaller for
rods than for spheres~\cite{toschi12}. The reason for this behavior is that the tendency to align with $\bm\omega$
is stronger for elongated particles~\cite{pumir11,toschi12} than for spheres.
In 2D, such an alignment cannot take place and
$\langle\dot{\theta}^2\rangle$ increases as $\gamma$ increases.

We have examined the statistics of the orientational and rotational dynamics
of elliptical tracers in 2D, homogeneous and isotropic turbulence.
By considering two sets of simulations with different $\kinj$, we have
shown that these properties depend on the scale at which the turbulent flow is
generated. In the small-$\kinj$ case,
the spatial correlation of the nematic order parameter indicates
the existence of large-scale structures in the spatial distribution of $\bm p$,
which are absent in the intermediate-$\kinj$ case. 
Moreover, the probability of $\bm p$ being aligned with $\nabla\times\bm
\omega$ or $\bm e_1$ is much lower for intermediate $\kinj$
than for small $\kinj$. These differences can be explained by noting
that the dynamics of fluid particles is different in the direct- and 
inverse-cascade regimes~\cite{BC00,BS02}, and hence the Lagrangian statistics
of $\nabla\bm u$ depends on $\kinj$ (see, \textit{e.g.},
$\tau_{S_{11}}$,  $\tau_{S_{12}}$, 
and $\tau_\omega$ given in Table~\ref{table:para}, as well
as the Lagrangian autocorrelation functions of the components of $\nabla\bm u$
reported in~\cite{SM}).
Our study sheds new light on the qualitative differences between 2D and
3D homogeneous, isotropic fluid turbulence. These differences lead
to a weaker alignment between $\bm p$ and $\nabla\times\bm
\omega$ in 2D as compared to the alignment between $\bm p$
and $\bm\omega$ in 3D and to a different dependence
of $\langle\dot{\theta}^2\rangle$ upon $\gamma$ (as $\gamma$ increases,
$\langle\dot{\theta}^2\rangle$ \textit{increases} in 2D
but \textit{decreases} in 3D).
We hope our comprehensive study of the statistical properties of
elliptical tracer particles in 2D, homogeneous and isotropic
turbulent fluid flows will stimulate experimental studies of
such particles.

\acknowledgments
We are grateful to G. Boffetta, D. Mitra, S. Musacchio, P.~Perlekar, and S.S. Ray for useful discussions.
We acknowledge support from the 
EU COST Action MP0806 ``Particles in Turbulence''
and the Indo--French Centre for Applied Mathematics (IFCAM). AG and RP thank UGC, CSIR, DST (India)
for support and SERC (IISc) for computational resources.


\begin{thebibliography}{10}

\bibitem{MY75}
A.S. Monin and A.M. Yaglom, \textit{Statistical Fluid Mechanics}
(Dover Publications, Inc., Mineola, NY, 1975).

\bibitem{frischbook}
U. Frisch, {\it Turbulence: The Legacy of A.N. Kolmogorov}
(Cambridge University Press, Cambridge, England, 1995).

\bibitem{BJPV98}
T. Bohr, M.H. Jensen, G. Paladin, and A. Vulpiani,
\textit{Dynamical Systems Approach to Turbulence}
(Cambridge University Press, Cambridge, England, 1998)

\bibitem{PPR09}
R. Pandit, P. Perlekar, and S.S. Ray, Pramana - Journal of Physics
\textbf{73}, 157 (2009).

\bibitem{FGV01}
G. Falkovich, K. Gaw\c{e}dzki, and M. Vergassola, Rev. Mod. Phys.
\textbf{73}, 913 (2001).

\bibitem{toschiannrev}
F. Toschi and E. Bodenschatz, Annu. Rev. Fluid Mech. {\bf 41}, 375 (2009).

\bibitem{collinsannrev}
J.P.L.C. Salazar and L.R. 
Collins, Annu. Rev. Fluid Mech.  {\bf 41}, 405 (2009).

\bibitem{KS11}
D.L. Koch and G. Subramanian, Annu. Rev. Fluid Mech. \textbf{43}, 637 (2011).

\bibitem{CL94}
J.P. Chen and D. Lamb, J. Atmos. Sci. \textbf{51}, 1206 (1994).

\bibitem{LSA11}
F. Lundell, L.D. S\"oderberg, and P.H. Alfredsson,
Annu. Rev. Fluid Mech. \textbf{43}, 195 (2011).

\bibitem{SK05}
E.S.G. Shin and D.L. Koch, J. Fluid Mech. \textbf{540}, 143 (2005).

\bibitem{pumir11}
A. Pumir and M. Wilkinson, New J. Phys. {\bf 13} 093030 (2011).

\bibitem{toschi12}
S. Parsa, E. Calzavarini, F. Toschi, and G. A. Voth, Phys. Rev. Lett.
{\bf 109}, 134501 (2012).

\bibitem{WK12}
M. Wilkinson and H.R. Kennard, J. Phys. A: Math. Theor. \textbf{45}, 455502 (2012).

\bibitem{V13}
D. Vincenzi, J. Fluid Mech. \textbf{719}, 465 (2013).

\bibitem{CM13}
L. Chevillard and C. Meneveau, J. Fluid Mech. \textbf{737}, 571 (2013).

\bibitem{GEM14}
K. Gustavsson, J. Einarsson, and B. Mehlig, Phys. Rev. Lett. \textbf{112}, 014501 (2014).

\bibitem{voth11}
S. Parsa, J.S. Guasto, M. Kishore, N.T. Ouellette, J.P. Gollub, and G.A. Voth,
Phys. Fluids {\bf 23}, 043302 (2011).

\bibitem{WBM09-10}
M. Wilkinson, V. Bezuglyy, and B. Mehlig, Phys. Fluids \textbf{21}, 043304
(2009);
V. Bezuglyy, B. Mehlig, and M. Wilkinson, Europhys. Lett. \textbf{89}, 34003
(2010)

\bibitem{L08}
M. Lesieur, \textit{Turbulence in Fluids} (Springer, Dordrecht,
The Netherlands, 2008).

\bibitem{BE12}
G. Boffetta and R.E. Ecke, Annu. Rev. Fluid Mech. \textbf{44}, 427 (2012).

\bibitem{PP10}
P. Perlekar and R. Pandit, New J. Phys. \textbf{12}, 023033 (2010) 
and references therein.

\bibitem{J22}
G.B. Jeffery, Proc. R. Soc. Lond. A \textbf{102}, 161 (1922).


\bibitem{canuto88}
C. Canuto, M. Y. Hussaini, A. Quarteroni, and T. A. Zang,
{\it Spectral Methods in Fluid Dynamics} (Springer-Verlag,
Berlin, 1988).

\bibitem{cox02}
S. M. Cox and P. C. Matthews, J. Comput. Phys. {\bf 176}, 430
(2002).

\bibitem{num_recp}
W. Press, B. Flannery, S. Teukolsky, and W. Vetterling,
{\it Numerical Recipes in Fortran} (Cambridge University
Press, Cambridge, 1992).

\bibitem{OW}
A. Okubo, Deep-Sea Res. Oceanogr. Abstr. \textbf{17}, 445 (1970);
J. Weiss, Physica (Amsterdam) \textbf{48D}, 273 (1991).

\bibitem{PRMP11}
P. Perlekar, S.S. Ray, D. Mitra, and R. Pandit, Phys. Rev. Lett.
\textbf{106}, 054501 (2011).

\bibitem{SM}
See Supplemental Material at URL for the
spatiotemporal evolution of the plots shown in Fig.~\ref{snapshot},
the PDFs of the alignment
conditioned on the sign of the Okubo--Weiss parameter, the PDFs
of $\dot{\theta}$ for different values of $\gamma$, and the Lagrangian
statistics of $\nabla\bm u$.

\bibitem{CL95}
P.M. Chaikin and T.C. Lubensky, \textit{Principles of Condensed Matter Physics}
(Cambridge University Press, Cambridge, England, 1995).

\bibitem{BM10}
G. Boffetta and S. Musacchio, Phys. Rev. E \textbf{82}, 016307 (2010).

\bibitem{BC00}
G. Boffetta and A. Celani, Physica (Amsterdam) \textbf{280A}, 1 (2000). 

\bibitem{BS02}
G. Boffetta and I.M. Sokolov, Phys. Fluids \textbf{14}, 3224 (2002).  

\end{thebibliography}
\end{document}


\title{Elliptical Tracers in Two-dimensional, Homogeneous, Isotropic
Fluid Turbulence: the Statistics of Alignment, Rotation, and Nematic Order}
\author{Anupam Gupta}
\affiliation{Center for Condensed Matter Theory,
Department of Physics, Indian Institute of Science, Bangalore 560012, India}
\author{Dario Vincenzi}
\affiliation{Univ. Nice Sophia Antipolis, CNRS, LJAD, UMR 7351,
06100 Nice, France}
\author{Rahul Pandit}
\affiliation{Center for Condensed Matter Theory,
Department of Physics, Indian Institute of Science, Bangalore 560012, India}

\maketitle

\begin{center}
\underline{\large\textit{Supplemental Material}}
\end{center}

\begin{figure}[h]
\centering
\subfigure[]{\includegraphics[width=0.28\columnwidth]{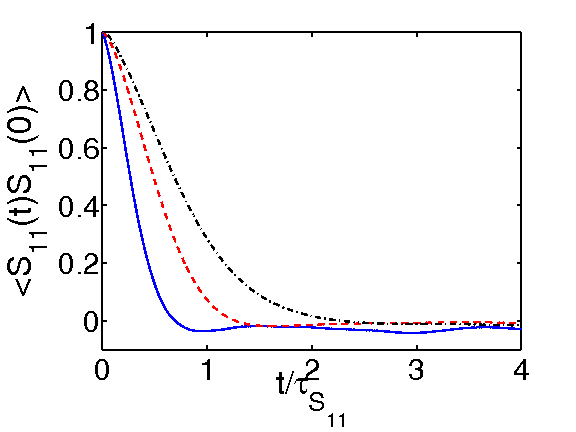}
\label{sufig:s11-k2}}%
\hspace{.5cm}%
\subfigure[]{\includegraphics[width=0.28\columnwidth]{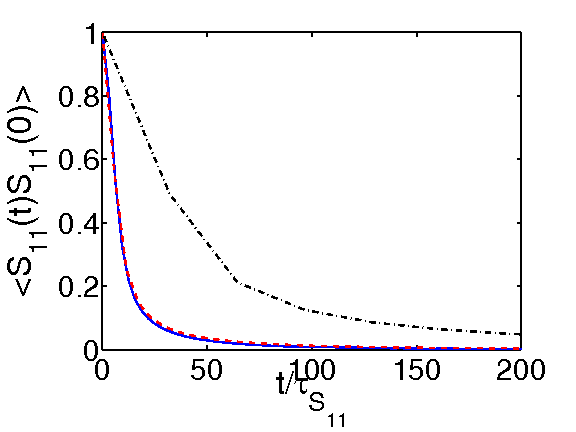}
\label{subfig:s11-k50}}
\\
\subfigure[]{\includegraphics[width=0.28\columnwidth]{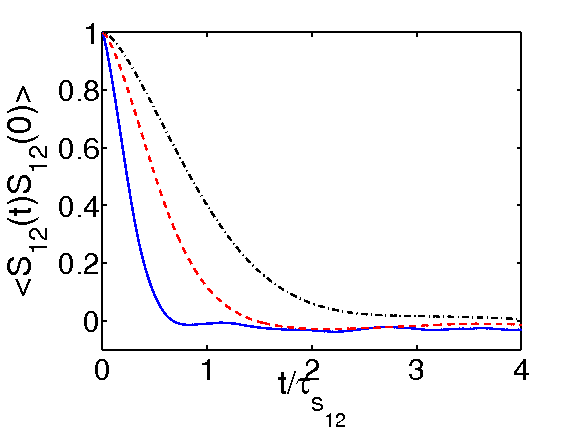}
\label{sufig:s12-k2}}%
\hspace{.5cm}%
\subfigure[]{\includegraphics[width=0.28\columnwidth]{c_s12_k50.eps}
\label{subfig:s12-k50}}
\caption{(Color online) Lagrangian autocorrelation functions of
the components of the rate-of-strain tensor
in case (A) (a,c) and in case (B) (b,d) for different values of 
$\mathit{Re}_\lambda$; the color code
is the same as in Fig.~2 in the main text.}
\label{fig:corr-strain}
\end{figure}

\begin{figure}[h]
\centering
\includegraphics[width=0.28\columnwidth]{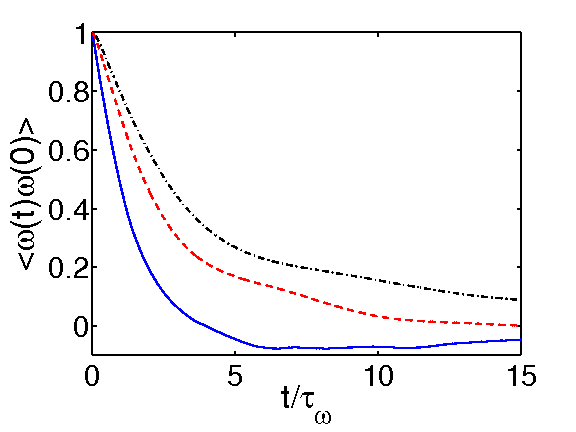}
\hspace{.5cm}%
\includegraphics[width=0.28\columnwidth]{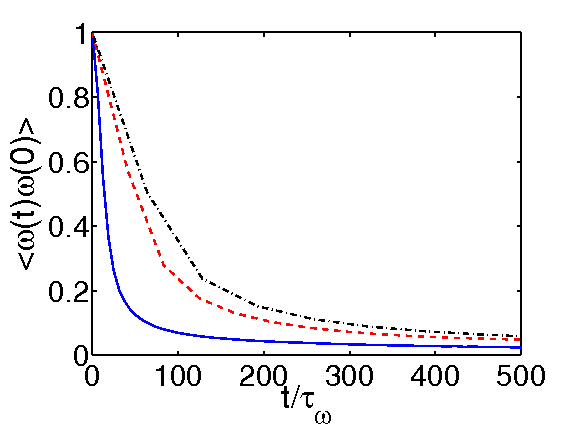}
\caption{(Color online) Lagrangian autocorrelation function of the vorticity
in case (A) (left) and in case (B) (right) for different values of 
$\mathit{Re}_\lambda$; the color code
is the same as in Fig.~2 in the main text.}
\label{fig:corr-vorticity}
\end{figure}

The Lagrangian autocorrelation functions
of the components of the rate-of-strain tensor and of the vorticity
are shown in Figs.~\ref{fig:corr-strain} and~\ref{fig:corr-vorticity}
for different values of $\mathit{Re}_\lambda$; 
the time variable is
rescaled by the appropriate correlation time.

\begin{figure}[h]
\centering
\includegraphics[width=0.28\columnwidth]{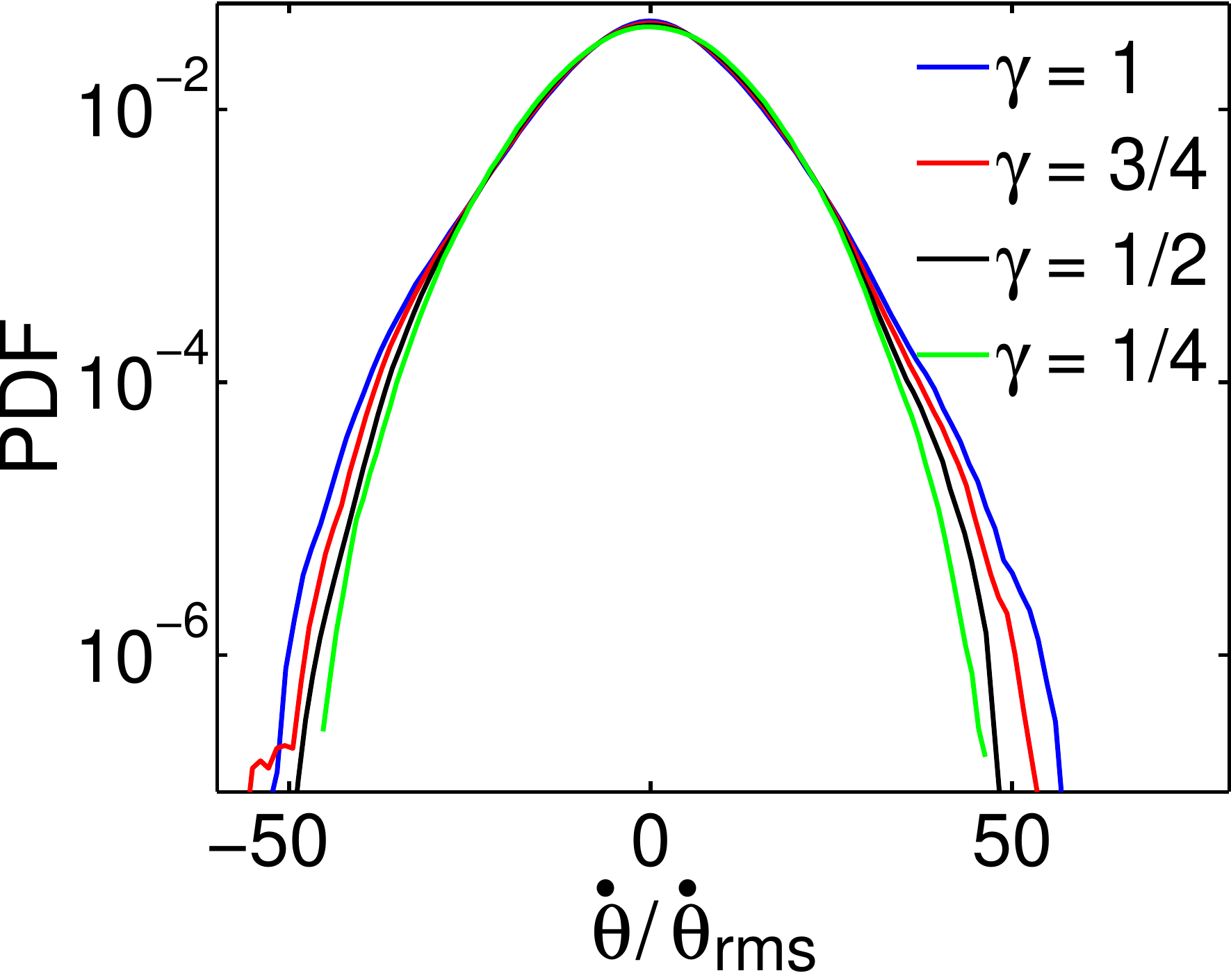}%
\hspace{.7cm}%
\includegraphics[width=0.28\columnwidth]{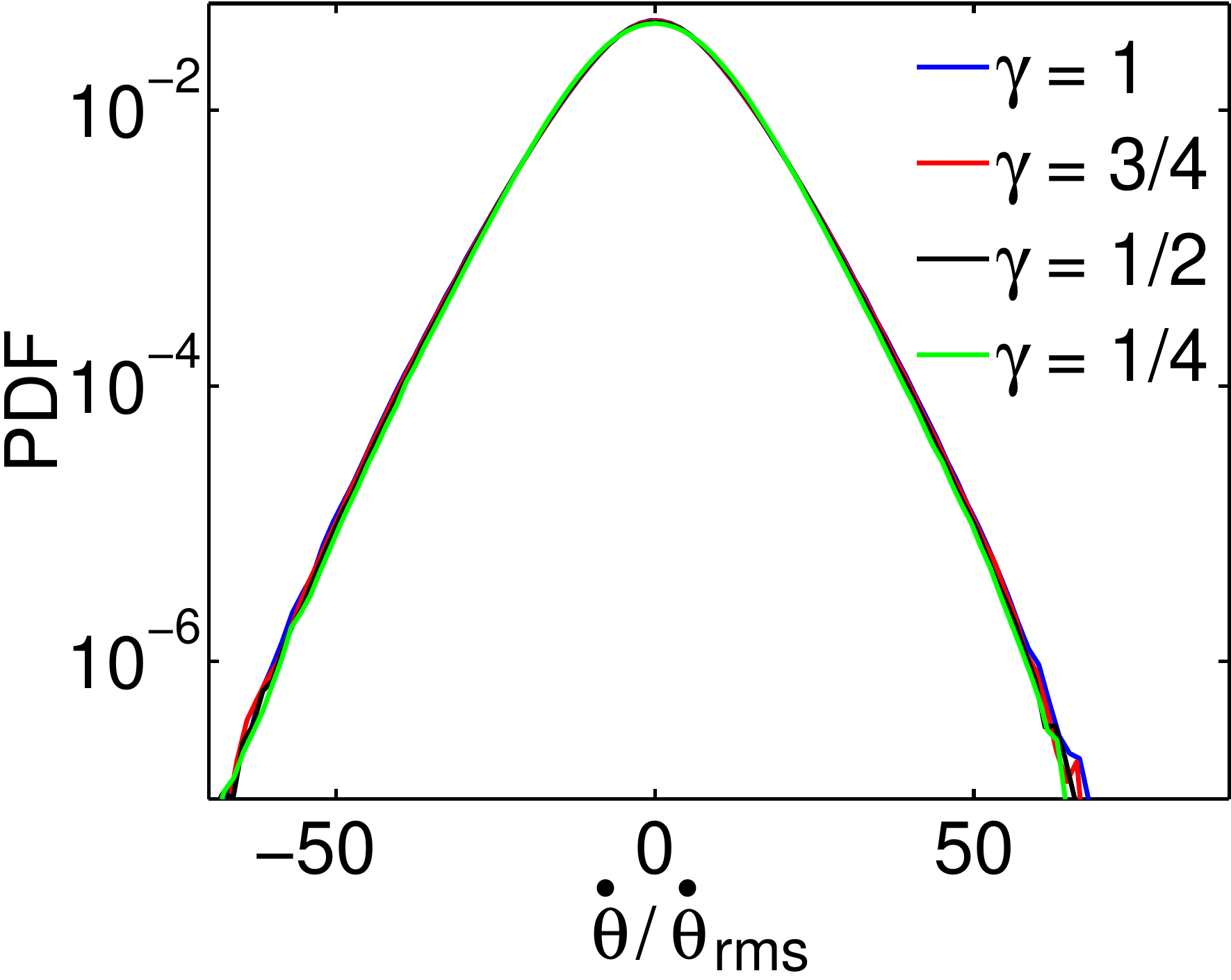}
\caption{(Color online) PDFs of the rotation rate
of elliptical tracer particles rescaled by their root-mean-square
values for different values of $\gamma$
for runs \texttt{A3} (left) and \texttt{B3} (right).}
\label{fig:pdf-rate-gamma}
\end{figure}

Figure~\ref{fig:pdf-rate-gamma} shows the PDFs of the rotation rate 
of particles for different values of $\gamma$.

\begin{figure}[h]
\centering
\includegraphics[width=0.28\columnwidth]{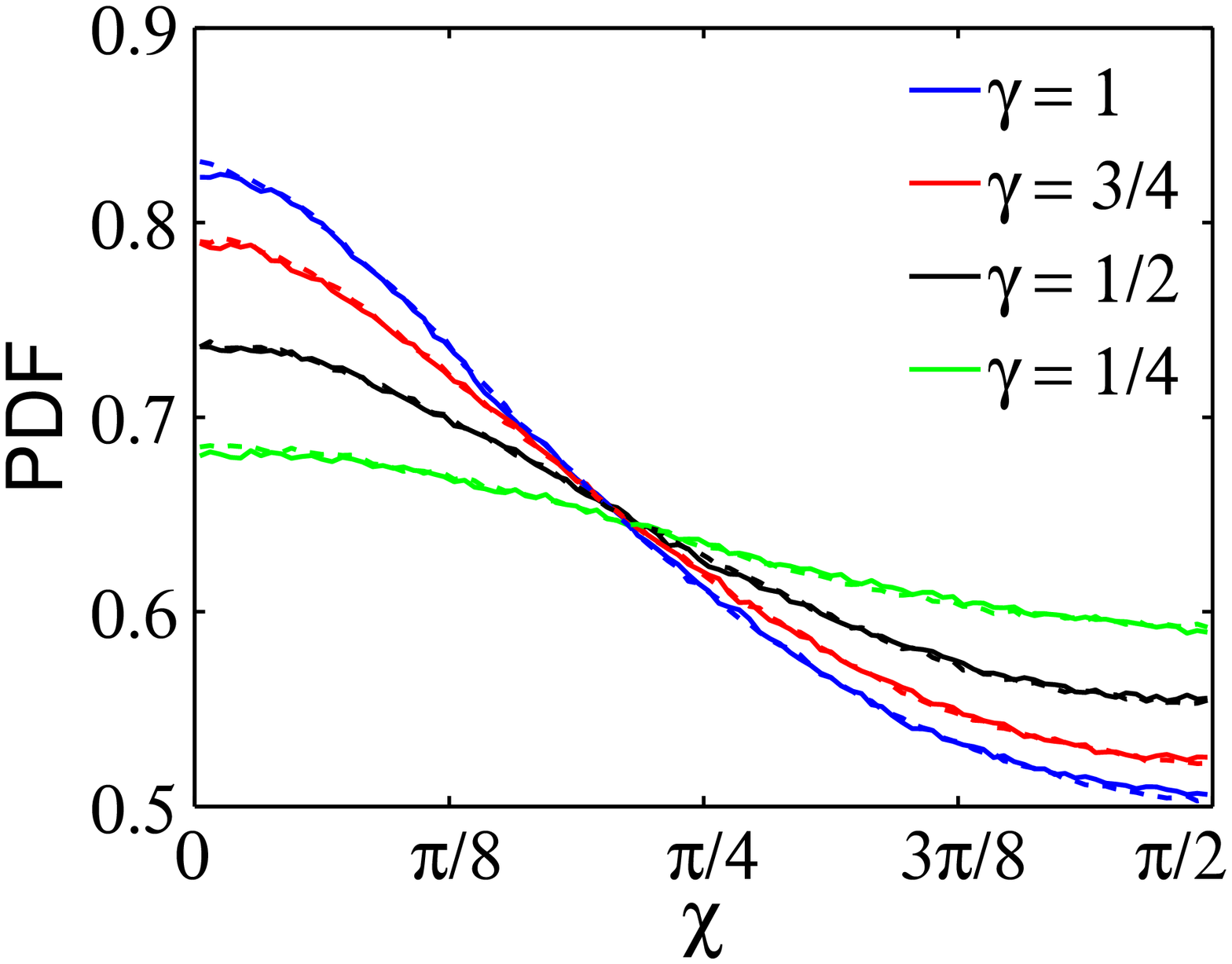}%
\hspace{.7cm}%
\includegraphics[width=0.28\columnwidth]{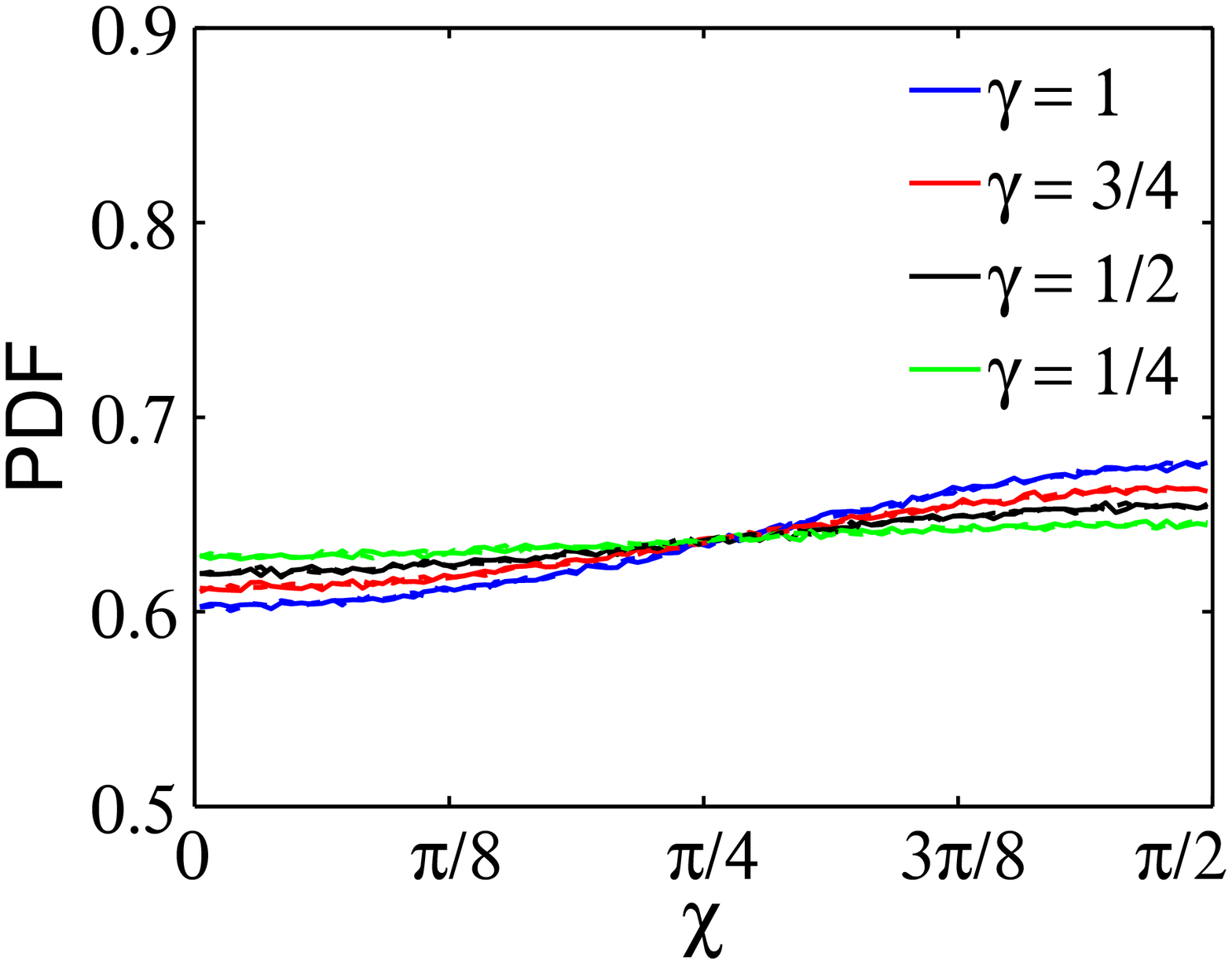}
\caption{(Color online)
Conditional PDFs of the angle $\chi$ on the sign of $\Lambda$ for different
 values of $\gamma$ for runs \texttt{A3} (left) and \texttt{B3} (right). 
The solid lines represent the conditional PDFs of $\chi$ given $\Lambda > 0$; 
the dashed lines represent the conditional PDFs of $\chi$ given $\Lambda < 0$.}
\label{fig:cond-curlw-p}
\end{figure}

\begin{figure}[h]
\centering
\includegraphics[width=0.28\columnwidth]{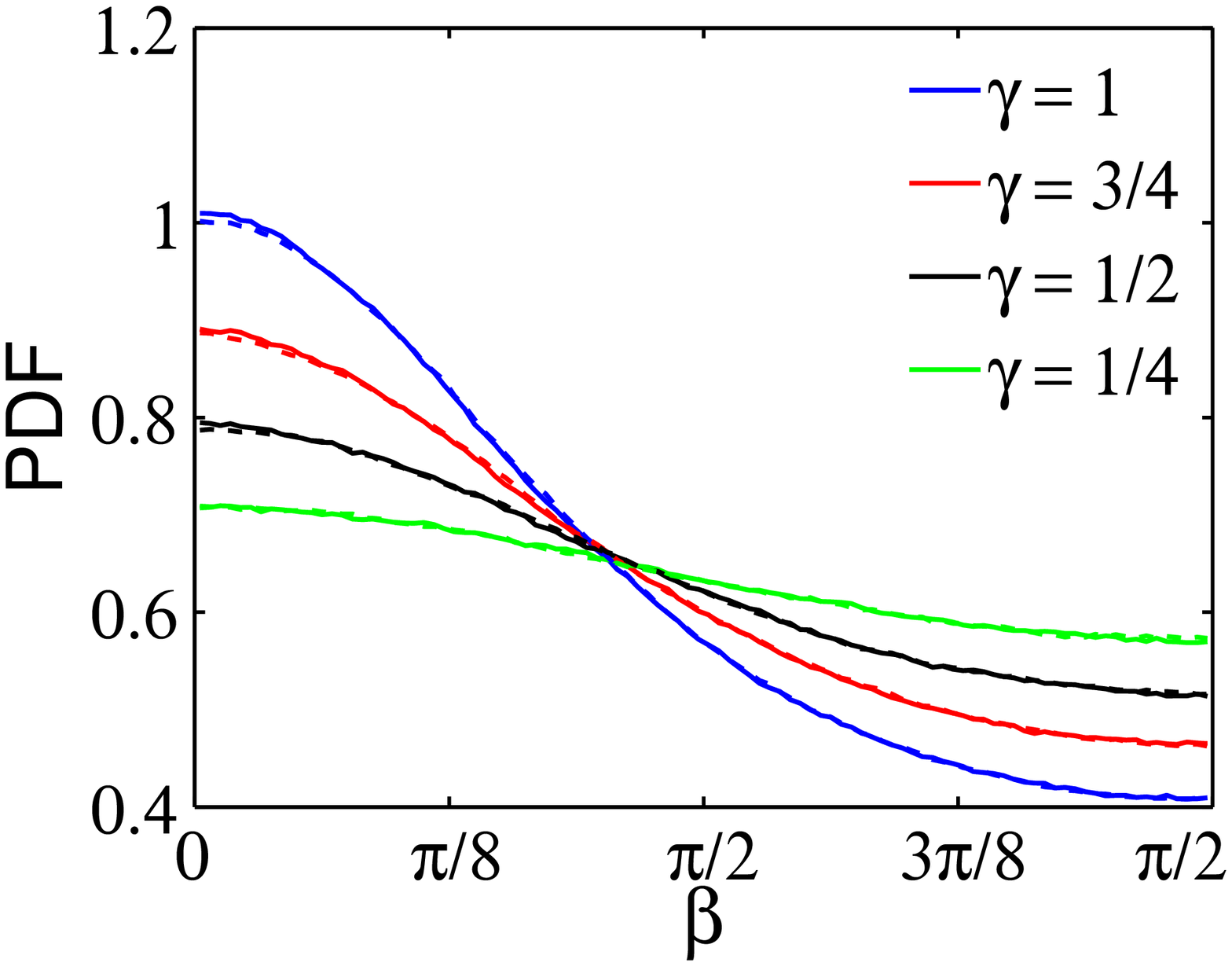}%
\hspace{.7cm}%
\includegraphics[width=0.28\columnwidth]{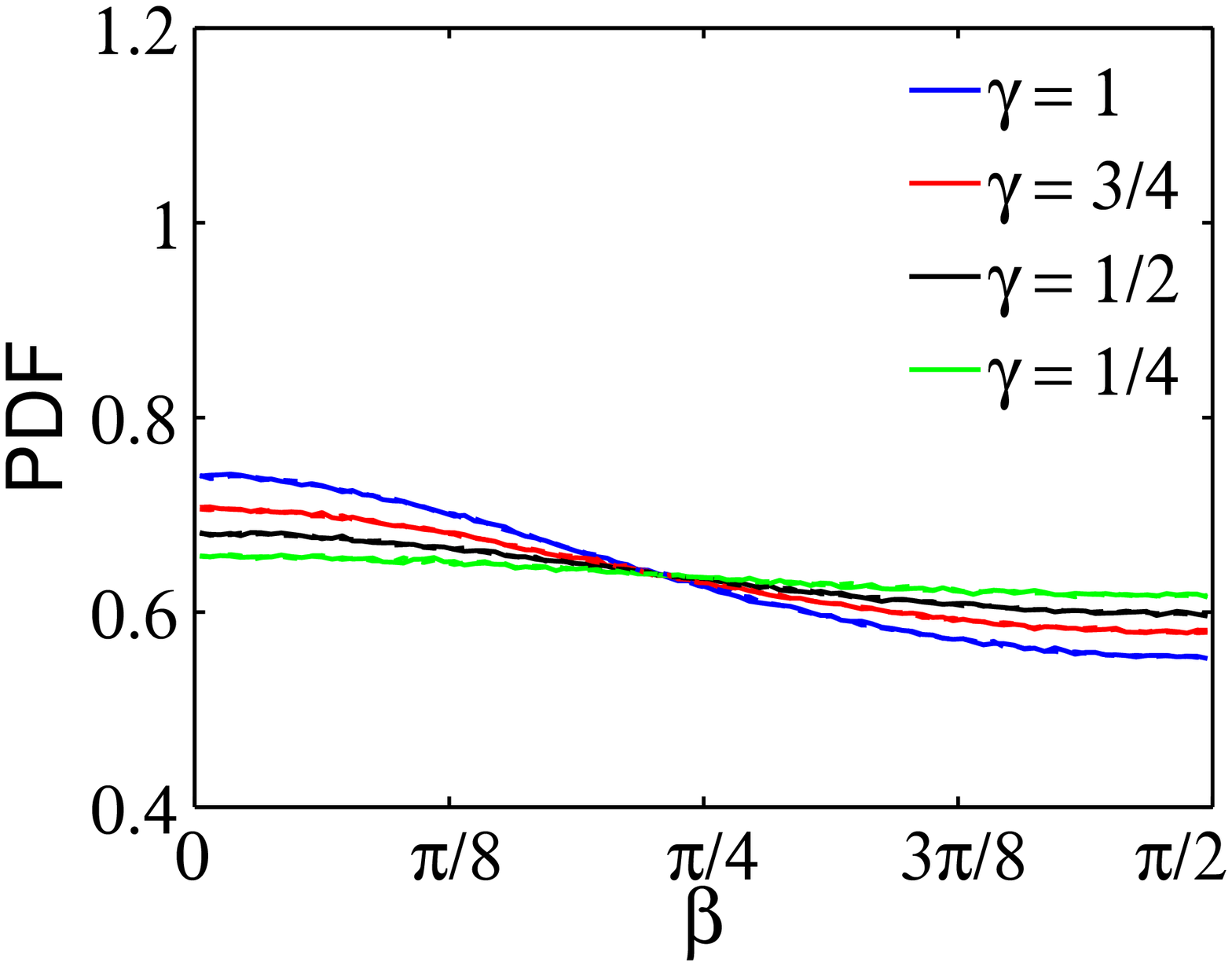}
\caption{(Color online) 
Conditional PDFs of the angle $\beta$ on the sign of $\Lambda$ for different
 values of $\gamma$ for runs \texttt{A3} (left) and \texttt{B3} (right). 
The solid lines represent the conditional of $\beta$ given $\Lambda > 0$; the dashed 
line represents the conditional PDFs of $\beta$ given $\Lambda < 0$.}
\label{fig:cond-e1-p}
\end{figure}

Figures~\ref{fig:cond-curlw-p} and \ref{fig:cond-e1-p} show the 
conditional PDFs of the angles $\beta$ and $\chi$ conditioned on the sign of the Okubo--Weiss 
parameter $\Lambda = (\omega^2-\sigma^2)/8$, where 
$\sigma^2 \equiv 4 \sum_{ij} S_{ij} S_{ij}$, 
for different values of $\gamma$.

